*Review*

# Contextuality or Nonlocality: What Would John Bell Choose Today?

Marian Kupczynski

Department of computer science and engineering, University of Quebec in Outaouais (UQO), Case Postale 1250, Succursale Hull, Gatineau, QC J8X 3X7, Canada; marian.kupczynski@uqo.ca

**Abstract:** A violation of Bell-CHSH inequalities does not justify speculations about quantum non-locality, conspiracy and retro-causation. Such speculations are rooted in a belief that setting dependence of hidden variables in a probabilistic model (called a violation of *measurement independence* (MI)) would mean a violation of experimenters' freedom of choice. This belief is unfounded because it is based on a questionable use of Bayes Theorem and on incorrect causal interpretation of conditional probabilities. In Bell-local realistic model, hidden variables describe only photonic beams created by a source, thus they cannot depend on randomly chosen experimental settings. However, if hidden variables describing measuring instruments are correctly incorporated into a contextual probabilistic model a violation of inequalities and an apparent violation of no-signaling reported in Bell tests can be explained without evoking quantum non-locality. Therefore, for us, a violation of Bell-CHSH inequalities proves only that hidden variables have to depend on settings confirming contextual character of quantum observables and an active role played by measuring instruments. Bell thought that he had to choose between non-locality and the violation of experimenters' freedom of choice. From two bad choices he chose non-locality. Today he would probably choose the violation of MI understood as *contextuality*.

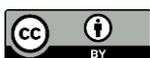

## 1. Introduction

Quantum mechanics (QM) provides probabilistic predictions and the main question debated since nearly 100 years is: are these probabilities irreducible or do they emerge from some more detailed description of physical reality and experiments used to probe it. Einstein strongly believed that QM should emerge from a more detailed description of individual physical systems [1, 2].

Bell was a realist who believed that physical objects possess definite properties [3,4]. In 1964, he proposed a probabilistic local realistic hidden variable model (LRHVM) trying to reproduce quantum predictions for an ideal EPRB experiment [3]. Pair-wise expectations deduced using LRHVM have to satisfy Clauser-Horne-Shimony–Holt inequalities (CHSH) [5, 6] which for some experimental settings are violated by quantum predictions and by experimental data in Bell Tests. The violation of inequalities is a source of unfounded speculations about the non-locality of Nature, free will and quantum magic.

In LRHVM, it is correctly assumed that hidden variables, describing only photonic beams created by a source, do not depend on randomly chosen experimental settings: $p(\lambda, x, y) = p(\lambda) p(x, y)$. Using the Bayes theorem, we obtain $p(\lambda | x, y) = p(\lambda)$ p($\lambda$ |x, y) = p ($\lambda$) and $p(x, y | \lambda) = p(x, y)$. This is why this assumption, called, *measurement independence* (MI), *free choice* or *no conspiracy* has been believed, for many years, to be a direct consequence of *experimenters' freedom of choice* (FC). For majority of scientists FC is a prerequisite of science and its violation would be unacceptable.

However, a choice of $(x, y)$ being labels of different experimental settings is followed by a choice of corresponding measuring instruments which, as fathers of QM taught us, are playing an active role in creation of measurement outcomes. Therefore, if one wants to explain outcomes of Bell Tests, using a hidden variable model, one should incorporate, in this model, variables describing measuring instruments. Of course, it does not guarantee that such an extended contextual hidden variable model (CHV) is successful.

In this article we review and generalize arguments given in [7–12], that a violation of MI (called more recently *statistical independence*) does not restrict FC. A misunderstanding is based on incorrect causal interpretation of conditional probabilities [9–12]. If $p(\lambda|x,y) \neq p(\lambda)$ then $p(x,y|\lambda) = p(\lambda|x,y) p(x,y) / p(\lambda) \neq p(x,y)$ but it does not mean, that $\lambda$ can causally influence, how $(x, y)$ are chosen. A *statistical dependence* does not imply a *causal dependence* and *correlation* does not mean *causation*.

CHV contains setting dependent (contextual) variables describing measuring instruments thus *statistical dependence* is incorporated into this model and easy to understand. As Bohr insisted one may not separate the behavior of atomic objects and the interaction with measuring instruments.

In QM experiments performed in incompatible experimental contexts are described by specific dedicated probabilistic models. In LRHVM, MI allows implementing random variables, describing random experiments performed using incompatible experimental settings, on a unique probability space, on which they are jointly distributed. Such implementation defines a *noncontextual probabilistic coupling*. In fact, CHSH are *noncontextuality inequalities* for a 4-cyclic Bell scenario [13]. Therefore, MI can be called *noncontextuality* and its violation *contextuality*.

*Statistical dependence* does not restrict FC, thus talking about a violation of *free choice* as a resource [10, 12] in Bell experiments is misleading. Of course, *statistical independence* could be violated due to *superdeterminism* understood as specific causal influences from the common past of photonic beams, experimenters and instruments they are using in distant laboratories. Next, hidden variables and experimenters' choices could be influenced in a way, to make experimental outcomes comply with QM. This explains why MI is called *free choice* or *no conspiracy* and Bell clearly preferred *non-locality* to the violation of *free choice*.

There are two probabilistic hidden variable models: LRHVM and a stochastic hidden variable model (SHVM) [5, 6]. Contrary to what several authors believe, LRHVM is not a special case of SHVM. These two probabilistic models allow deriving CHSH inequalities, but they describe completely different random experiments requiring different experimental protocols. In LRHVM for a fixed $\lambda$ distant outcomes are strictly predetermined and correlated. In SHVM for a fixed $\lambda$ they are created in two random independent experiments. In LRHVM, 'entangled photon pairs' are described as *pairs of socks*. In SHVM, they are described as *pairs of dice* and in each setting $(x, y)$ we have a different family of stochastically independent distant random experiments labeled by $\lambda$. A detailed discussion of these two models and experimental protocols, implied by them, may be found in [7].

In LRHVM, outcomes (clicks on detectors coded ±1) are locally predetermined by variables describing correlated photonic signals, produced by a source. Local predetermination of outcomes of all experiments, by some ontic properties of signals, is called usually: *local realism, classicality* or *counterfactual definiteness*. *Local realism* implies MI and the existence of a *probabilistic coupling*. CHSH are significantly violated in various Bell Tests. Therefore, experimental outcomes are not predetermined by the properties of photonic signals and as Peres correctly concluded: *unperformed experiments have no results* [14].

Various metaphysical assumptions may motivate a choice of a probabilistic model. However, once a model is chosen, its meaning and its implications can only be rigorously discussed in a probabilistic framework. This is why, for us, saying that $\lambda$ can be anything even the quantum wave function of Universe is misleading and meaningless. Hidden random variables describe details of an experimental protocol consistent with a given probabilistic model [7].

The ideal EPRB experiments and perfect correlations do not exist [15]. Random variables describing data in Bell Tests are inconsistently connected and should be analyzed using Contextuality-by-Default approach (CbD) [11, 16–23]. In CbD, proposed by Dzhafarov and Kujala, all empirical scenarios are described by systems of random variables representing measurements of properties $q$ in contexts $c$. Properties of experimental scenarios and possible hidden variable models are studied without evoking any metaphysical assumptions. *Free choice* is equivalent to context-independent mapping and experimenters' *free will* assumption is completely redundant [21–23]. In this paper, we are not using CbD approach. We define and explain only specific probabilistic couplings using a simplified notation.

At the end of this introduction, we want to summarize in a simpler language, what is the main motivation of this paper? What is new, what the conclusions are and how they are derived in detail in subsequent sections.

It is well known that CHSH may be derived for 4 jointly distributed random variables describing a random experiment in which 4 outcomes are outputted in each trial. In such experiment estimates of pair-wise expectations obey this inequality for any finite sample.

It is clear, that random variables describing outcomes of EPRB and Bell Tests are not jointly distributed and Bell never claimed the opposite. One may ask: how could he derive his inequalities. He could do it, because in LRHVM, there exists a JP of 4 random variables, and he used underline{implicitly} this JP in his proof. In the literature about Bell inequalities no distinction is made between these two sets of random variables. In Section 2 we decided to make a distinction.

It allows explaining more clearly, how LRHVM defines a probabilistic coupling for EPRB experiments. This probabilistic coupling is neither consistent with quantum predictions nor with experimental data in Bell Tests.

The existence of a probabilistic coupling does not mean that CHSH hold in the experiments performed using incompatible settings and described using this coupling. CHSH can be and are violated in these experiments by estimates of pair-wise expectations obtained using finite samples. The existence of a probabilistic coupling only allows deriving probability bounds on how large and how frequent violations may be observed in these experiments. This problem is discussed in Section 3.

In Section 4 we explain a violation of inequalities and an apparent violation of non-signaling, reported in Bell Tests, by incorporating into a probabilistic model, variables describing measuring instruments. If $(i, j)$ $(i, j)$ denote 4 incompatible settings in Bell Tests, using Bayes Theorem, we demonstrate that $p(i, j \mid \lambda) = 1$, what only means: if a 'hidden event' $\{(\lambda_1, \lambda_2, \lambda_i, \lambda_j)\}$ 'happened', then the settings $(i, j)$ were used. Thus, the violation of *statistical independence* in our model does not justify speculations about the violation of *free choice* or *conspiracy*. This result was derived for the first time in [9] and reproduced in [11, 12]. In previous papers [9–12] we assumed that $p_{ij}(\lambda_i, \lambda_j) = p_i(\lambda_i) p_j(\lambda_j)$. In Section 4, we prove, that our conclusions hold even, if $p_{ij}(\lambda_i, \lambda_j)$ do not factorize.

In Section 5 we rectify some misleading claims of Lambare and Franco, made in [24]. In particular we reject their criticism of correct arguments given in [25–33].

Section 6 contains some final conclusions and additional discussion of *local realism*, *statistical independence* and *superdeterminism*.

## 2. LRHVM and Probabilistic Coupling

We discuss LRHVM and its implications using a rigorous probabilistic framework, what avoids misunderstanding.

The experimental protocol of an ideal EPRB is the following [15]:

1. A beam (ensemble) E of entangled pairs of particles is created by a source. One particle is sent to Alice and its twin partner to Bob in distant laboratories, who chose independently experimental settings $(x, y)$ of their polarization beam splitters (PBS). In general $(x, y)$ are labels and not necessarily values of some random variables.
2. Particles pass by corresponding beam splitters (PBS) and produce clicks on detectors, which are coded by two random variables $A_x$ and $B_y$ taking values ±1.

In QM, it does not matter, how settings $(x, y)$ are chosen. Experiments performed using incompatible settings are described, by specific, setting dependent, probability distributions. In particular pair-wise expectations for a setting $(x, y)$ are given by:

$$E(A_x B_y) = Tr \rho \hat{A}_x \hat{B}_y \qquad (1)$$

where $\rho$ is a density matrix describing the ensemble E prepared by a source, $\hat{A}_x$ and $\hat{B}_y$ are operators representing spin projection measurements made by Alice and Bob, respectively.

As Cetto et al. pointed out in [34], the Equation (1) can be rewritten as:

$$E(A_x B_y) = \sum_{a,b} ab \, p_{xy}(a,b) \qquad (2)$$

where $a = \pm 1$ and $b = \pm 1$ are experimental outcomes being eigenvalues of the operators $\hat{A}_x$ and $\hat{B}_y$. The quantum probabilistic models (1) and (2) explicitly depend on settings. If settings were changed, then the quantum description would change.

For a singlet state $\rho$ and for identical settings $(x, x)$ QM predicts $p(A_x = 1) = 1/2$, $p(B_y = 1) = 1/2$ and $p(A_x = 1, B_x = -1) = 1$. It is mind boggling, <u>if one believes</u> that quantum randomness is perfect and irreducible. Perfect randomness, by definition, is incompatible with strictly correlated outcomes [33]. The outcomes of flipping a fair coin cannot be predicted, thus when two coins are flipped their outcomes cannot be <u>always</u> strictly correlated. It is believed that quantum randomness is perfect and irreducible. However, it is difficult to prove it because subsequent digits in the decimal approximation of a number $\pi$ pass with success all randomness tests even, if they are strictly determined. As a matter of fact, the violation of inequalities in Bell Tests gave not only the arguments against LRHVM but also against SHVM and irreducible randomness.

This is why Bell assumed that experimental outcomes in an ideal EPRB are predetermined by correlated properties of particles prepared at the source. An apparent randomness and a statistical scatter of outcomes are then due, similar as in classical physics, to a lack of knowledge of the statistical ensemble E.

Let us cite Bell [35]: *"For me, it is so reasonable to assume that the photons in those experiments carry with them programs, which have been correlated in advance, telling them how to* *behave. This is so rational that I think that when Einstein saw that, and the others refused to see it, he was the rational man."*

It is well known that CHSH inequalities may be derived for a random experiment in which 4 outcomes are outputted in each trial and described by 4 jointly distributed random variables on a unique probability space. It is obvious that the random variables $(A_x, B_y, A_{x'}, B_{y'})$ are not jointly distributed and Bell never claimed the opposite. Nevertheless, in LRHVM he used implicitly a joint probability of 4 random variables in order to prove the inequalities. We explain below that he, in fact, postulated the existence of a probabilistic coupling. In order to make it easier to understand we are using primed random variables $(A'_x, B'_y, A'_{x'}, B'_{y'})$, which in LRHVM are jointly distributed. Bell did not make this distinction, but we do and we define LRHVM as below:

$$E(A'_x B'_y) = \sum_{\lambda \in \Lambda} A_x(\lambda) B_y(\lambda) p(\lambda) \tag{3}$$

Please note, that Bell replaced $E(A'_x B'_y)$ in the Formula (3) by $E(A_x B_y)$. We use primed random because, there is no JP of $(A_x, B_y, A_{x'}, B_{y'})$, but there exists a JP of primed variables $(A'_x, B'_y, A'_{x'}, B'_{y'})$. Namely for four experimental settings $(x, y) = (i, j) = (1, 1), (1, 2), (2, 1)$ or $(2, 2)$ we have:

$$E(A'_1 B'_1 A'_2 B'_2) = \sum_{\lambda \in \Lambda} A_1(\lambda) B_1(\lambda) A_2(\lambda) B_2(\lambda) p(\lambda) \tag{4}$$

Moreover, there exists a mapping $M : \Lambda \Rightarrow \Omega = \{\omega = (a_1, b_1, a_2, b_2)\}$, where $a_i = A_i(\lambda) = \pm 1$ and $b_j = B_j(\lambda) = \pm 1$ thus:

$$E(A'_1 B'_1 A'_2 B'_2) = \sum_{\omega \in \Omega} a_1 b_1 a_2 b_2 \, p(a_1, b_1, a_2, b_2) \tag{5}$$

and instead of (3) we may use:

$$E(A'_i B'_j) = \sum_{\omega \in \Omega_{ij}} a_i b_j \, p_{A'_i B'_j}(a_i, b_j) \tag{6}$$

where $\Omega_{ij} = \{(a_i, b_j)\}$ and $p_{A'_i B'_j}(a_i, b_j)$ is a standard marginal distribution obtained from $p(a_1, b_1, a_2, b_2) = p(\omega) = \sum_{\lambda \in M^{-1}(\omega)} p(\lambda)$. Please note, that sample space $\Omega$ contains exactly 16 elements and each sample space $\Omega_{ij}$ only 4 elements. Using (5) and (6) one easily obtains CHSH inequalities [4]:

$$|E(A'_1 B'_1) + E(A'_1 B'_2) + E(A'_2 B'_1) - E(A'_2 B'_2)| \leq 2 \tag{7}$$

As Fine demonstrated [36, 37], the inequalities (7) are necessary and sufficient conditions for the existence of JP defined above.

In EPRB such JP does not exist. Nevertheless, Bell postulated from the beginning that:

$$E(A_i B_j) = E(A'_i B'_j) = \sum_{\lambda \in \Lambda} A_i(\lambda) B_j(\lambda) p(\lambda) \tag{8}$$

without noticing that his proofs implicitly rely on the existence of a counterfactual JP [3, 4]. He demonstrated that, for some experimental settings, the inequalities were violated by quantum predictions (1-2), but in 1964, he still hoped that experimental data might agree with his model.

In CbD [16–23], the Equation (8) defines a non-contextual coupling of only pairwise jointly measurable observables:

$$P(A_i = a) = P(A'_i = a); P(B_j = b) = P(B'_j = b); E(A_i B_j) = \langle A_i B_j \rangle = \langle A'_i B'_j \rangle \quad (9)$$

which in general does not exist. To indicate explicitly, that the experiments performed in different settings are incompatible, in CbD one would replace in (8-9) $A_i$ by $A_{ij}$ and $B_j$ by $B_{ij}$. Since in EPRB *no-signaling* is not violated, thus we used a simplified notation.

Ideal EPRB experiments, with perfectly correlated clicks on distant detectors, do not exist [15, 33]. Nevertheless, a significant violation of (7) was reported in several Bell Tests. Thus, the data in these experiments can neither be described using LRHVM nor by SHVM.

For a mathematician, the violation of (7) means only, that a non-contextual probabilistic coupling (9) does not exist and that CHSH inequalities are simply *noncontextuality inequalities* for a 4-cyclic scenario [13] which can be rigorously derived for a random experiment such that in each trial 4 experimental outcomes $(a_1, b_1, a_2, b_2)$ are outputted.

## 3. Experimental Protocols and Finite Samples

Probabilistic models describe a scatter of observed outcomes without entering into details how these data were produced. However, there is an intimate relation between probabilistic models and experimental protocols [7, 38]. If we assume, that experimental settings are randomly chosen for each successive trial as it is carried out in Bell Tests, the model (3) describes a three-step random experiment.

1. A marble is drawn from an urn (or a box) E. Properties of marbles in E are described by $\lambda$ being values of a random variable $L$ distributed according to a probability distribution $p(\lambda)$ on a unique probability space $\Lambda$.
2. Experimenters, choose at random one among 4 available incompatible settings $(i, j)$ of their instruments, which output two numbers $a_i = A_i(\lambda)$ and $b_j = B_j(\lambda)$.
3. The marble is returned to the box and another marble is drawn from the box.

Since $A'_i = A_i(L)$ and $B'_j = B_j(L)$, there exists a JP of all these random variables. It is obvious, that the random variable $L$ and its probability distribution do not depend on how the settings $(i, j)$ are chosen in the step 2 of the experimental protocol. As in QM, $(i, j)$ are only labels of 4 incompatible experimental settings and *experimenters' freedom of choice* (FC) is never compromised.

In Bell Tests, instead of a marble we have 'pairs of photons'. In LRHVM they are described as pairs of socks which may have different colors and sizes. In SHVM, they are described as pairs of dice. More detailed discussion of these probabilistic models and their intimate relation with experimental protocols may be found in [7].

In LRHVM, each experiment $(i, j)$ is described as a fair sampling from $\Lambda$ followed by a deterministic assignment of outcomes $(A_i(\lambda), B_j(\lambda))$. If we limit ourselves to 4 settings, then as we saw in (5, 6), instead of $\Lambda$, we may use a finite sample space containing only 16 elements: $\Omega = \{a_1, b_1, a_2, b_2\}$, where $a_i = \pm 1$ and $b_j = \pm 1$. For each experimental setting $(i, j)$, in each trial, only two outcomes $(a_i, b_j)$ are outputted. If we estimate expectations $E(A'_i B'_j)$ using finite samples of size N these estimated expectations violate the inequalities (7) approximately 50% of time [15, 39–42], but not as significantly as predicted by QM and reported in Bell Tests.

An experimental protocol consistent with LRHVM is similar to the experimental protocol of a following thought experiment [39]. In each trial $(a_1, b_1, a_2, b_2)$ is drawn from Ω and displayed as a line in 4N × 4 spreadsheet. Next, setting labels $(i, j)$ are randomly chosen and outcomes $(a_i, b_j)$ are outputted and displayed in a corresponding N×2 spreadsheet and another quadruplet is drawn from Ω. If by chance, each pair of settings is chosen N times at the end we have four N × 2 spreadsheets, which may be used to estimate expectations $E(A'_i B'_j)$ and checking (7). We see that in each trial of this thought experiment outcomes are predetermined and measuring instruments passively register corresponding predetermined values.

In real experiments, 4N × 4 spreadsheets do not exist and four N × 2 spreadsheets describing the data obtained using 4 incompatible settings, are not simple random samples drawn from columns of some 4N × 4 spreadsheet [8,33]. They cannot be reordered, what was claimed in [24] to satisfy (7) and the only constraint, without additional assumptions, on estimated $E(A_i B_j) \neq E(A'_i B'_j)$, is: S ≤ 4 [11].

## 4. Violation of Statistical Independence in Bell Tests

In Bell Tests, some data violate no-signaling [41–45], thus they are also inconsistent with quantum predictions (1-2) for an ideal EPRB. Using CbD terminology [16–23], the data used to estimate pair-wise expectations are described by inconsistently connected random variables, thus they should be analyzed using CbD approach [11,12].

It is clear, that LRHVM and SHVM are oversimplified probabilistic models unable to describe these experimental data from Bell Tests. As Theo Nieuwenhuizen [46–48] correctly concluded, LRHVM suffers from *contextuality loophole*, because it does not incorporate correctly hidden variables describing measuring instruments, as they are perceived by incoming photonic signals.

If setting dependent hidden variables, describing instruments, are added to LRHVM, the data in Bell Tests may be described by a contextual probabilistic model:

$$E(A_{ij} B_{ij}) = \sum_{\lambda \in \Lambda_{ij}} \tilde{A}_i(\lambda_1, \lambda_i) \tilde{B}_j(\lambda_2, \lambda_j) p_{ij}(\lambda) \qquad (10)$$

where $\tilde{A}(\lambda_1, \lambda_i) = \pm 1; \tilde{B}_j(\lambda_2, \lambda_j) = \pm 1$, $\lambda = (\lambda_1, \lambda_2, \lambda_i, \lambda_j)$, $\Lambda_{ij} \cap \Lambda_{i'j'} = \emptyset$ and

$$p_{ij}(\lambda) = p(\lambda | i, j) = p_{ij}(\lambda_i, \lambda_j) p(\lambda_1, \lambda_2) \qquad (11)$$

This model violates statistical independence and $p(i, j | \lambda) \neq p(i, j)$ but, contrary to what is often claimed, it does not give arguments in favor of *superdeterminism*.

Using (11) and Bayes Theorem we obtain

$$p(\lambda, i, j) = p_{ij}(\lambda_i, \lambda_j) p(\lambda_1, \lambda_2) p(i, j) = p(\lambda) \rightarrow p(i, j | \lambda) = 1 \qquad (12)$$

The equation: $p(i, j | \lambda) = 1$ tells only, that if a hidden 'event' $\{(\lambda_1, \lambda_2, \lambda_i, \lambda_j)\}$ 'happened' then the settings $(i, j)$ were used [9–12]. It has nothing to do with *conspiracy* and FC is not compromised. In each trial, labels $(i, j)$ of experimental settings are chosen in two distant random experiments which do not depend how photonic beams are produced and how they are going to be processed later in the experiment.

However, if an event $\{(i, j)\}$ occurred then specific instruments described by $(\lambda_i, \lambda_j)$ are used. It was explained in detail for the first time in [9]. Therefore, the as-sumption $p(\lambda|i,j) \neq p(\lambda)$ may be called Bohr *contextuality* and not *conspiracy* or *superdeterminism*. In CbD approach, *contextuality* has a different more restricted meaning.

In (11), variables describing photonic signals are causally independent but stochastically dependent; variables describing measuring instruments are also causally independent but stochastically dependent $p_{ij}(\lambda_i, \lambda_j) \neq p_i(\lambda_i) p_j(\lambda_j)$. Variables $(i, j)$ are causally and stochastically independent from $(\lambda_1, \lambda_2)$.

In Bell Tests, two distant time series of clicks have to be converted into two discrete correlated samples using synchronized time windows. In [9, 11, 15, 33], we described raw data in Bell Tests using a model similar to (10), but with $A_i(\lambda_1, \lambda_i) = \pm 1, 0$, $B_j(\lambda_2, \lambda_j) = \pm 1, 0$ and $p_{ij}(\lambda_i, \lambda_j) = p_i(\lambda_i) p_j(\lambda_j)$. By conditioning on pairs of non-vanishing outcomes we derived a model describing the final set of data violating CHSH and no-signaling. Setting dependent pairing of distant outcomes was the origin of stochastic dependence of $(\lambda_i, \lambda_j)$ in (10).

The model (10) allows explaining data from Bell Tests in a local and causal way. Choosing a particular *statistical dependence* more precise predictions for expectations $E(A_{ij}B_{ij})$ may be made [49]. We know that transmission probabilities between two polarization filters obey the Malus law, which depends only on $cos(\theta)$, where $\theta$ is a relative angle between polarization axes of polarization filters. Therefore, if $\lambda$ are hidden variables describing a polarization filter how it is 'perceived' by an incoming beam at the moment of a measurement, it is plausible to assume, that after a rotation by an angle $\theta$, the same filter is described by variables $\lambda' = f(\lambda, cos(\theta))$. Therefore, one may try to explain $\theta$ dependence of estimated expectations $E(A_{ij}B_{ij})$, assuming that $p_{ij}(\lambda_i, \lambda_j)$ is also a function of $cos(\theta_{ij})$. The importance of rotational invariance was strongly advocated by Karl Hess [50].

## 5. Fine's Theorem and Joint Probabilities

In [24], Lambare and Franco made some statements, which we want to rectify.

They claim that talking about joint probabilities is misleading because LRHVM, which they call LHV model, may be derived using local causality, perfect correlations and MI. However, they do not realize that the conjunction of these three assumptions implies the existence of a counterfactual non-contextual probabilistic coupling (9) and a joint probability of 4 random variables (JP), which was implicitly used by Bell to derive CHSH. The authors' finite sample proof of CHSH fails, if JP does not exist. If MI is violated $p(\lambda|x, y) \neq p(\lambda)$ then JP does not exist and, for different settings, sampling is made from different probability spaces. Lambare and Franco dismissed the violation of *statistical independence*, believing that the violation of MI would mean *conspiracy*.

Fine demonstrated, that CHSH are necessary and sufficient conditions for the existence of JP of 4 only pair-wise measurable random variables [35, 36]. Nobody claims that Fine has disproved Bell's Theorem. Bell's Theorem is a mathematical theorem which says: *if LRHVM is used to describe EPRB, then some pair-wise cyclic expectations obey Bell-CHSH inequalities, which for some settings are violated by quantum predictions*. JP of n-random variables only exists, if in each trial n-results are outputted [7, 9, 38, 51]. Therefore, JP neither exists in Bell scenario nor in their counterexample [24].

In their counterexample we have: 7 random variables: $L$ taking values $\lambda \in \Lambda = \{1, 2, 3, 4, 5, 6\}$, $X$ taking values x = {1, −1}, $Y$ taking values y = {1, −1}, $A_x$ and $B_y$. $L$ describes an experiment in which hidden variables are sampled from $\Lambda$ by rolling a dice, $X$ and $Y$ are random variables describing flipping fair coins in order to determine experimental settings $(x, y)$, $A_x = A(x, L)$ and $B_y = B(y, L)$ are random variables describing predetermined outcomes. Namely:

$$A_x(\lambda) = A(x, \lambda) = x^\lambda, B_y(\lambda) = B(y, \lambda) = y^{\lambda+1} \tag{13}$$

We have 4 incompatible experiments, labeled by $(x, y)$, and only 2 outcomes are outputted in each trial, thus JP of 4 random variables $(A_1, A_{-1}, B_1, B_{-1})$ <u>does not exist</u>. It is easy to evaluate 4 expectations entering the inequality (7):

$$E(A_1 B_1) = 1, E(A_1 B_{-1}) = E(A_{-1} B_1) = 0, E(A_{-1} B_{-1}) = -1 \tag{14}$$

Expectations (14) do not violate the inequality (7). Lambare and Franco incorrectly conclude: "according to Fine's theorem A, a joint probability $P(A_1; A_{-1}; B_1; B_{-1})$ exists, although the experiments are incompatible".

The random variables $(A_1, A_{-1}, B_1, B_{-1})$ are not jointly distributed. Nevertheless, there exist 4 jointly distributed random variables $(A'_1, A'_{-1}, B'_1, B'_{-1})$, which define a non-contextual coupling $E(A_x) = E(A'_x), E(B_y) = E(B'_y), E(A_x B_y) = E(A'_x B'_y)$.

Instead of (4) using (13) we have now:

$$E(A'_1 A'_{-1} B'_1 B'_{-1}) = \sum_{\lambda \in \Lambda} A_1(\lambda) A_{-1}(\lambda) B_1(\lambda) B_{-1}(\lambda) p(\lambda) = \sum_{\lambda=1}^{6} 1^\lambda (-1)^\lambda 1^{\lambda+1} (-1)^{\lambda+1} \frac{1}{6} = -1 \tag{15}$$

The random variables $(A'_1, A'_{-1}, B'_1, B'_{-1})$ define a mapping $M: \Lambda \Rightarrow \Omega = \{(1,1,1,-1), (1,-1,1,1)\}$ and their joint probability distribution is:

$$p_1 = p(1,1,1,-1) = \frac{1}{2}, \quad p_2 = p(1,-1,1,1) = \frac{1}{2} \tag{16}$$

Using (16) we immediately obtain:

$$E(A'_1 A'_{-1} B'_1 B'_{-1}) = 1 \times 1 \times 1 \times (-1) \times p_1 + 1 \times (-1) \times 1 \times 1 \times p_2 = -1 \tag{17}$$

$E(A'_{-1} B'_1) = E(A'_1 B'_{-1}) = 0$ and $E(A'_{-1} B'_{-1}) = -1$. We use only (16) and we do not need to mention hidden variables.

Jointly distributed $(A'_1, A'_{-1}, B'_1, B'_{-1})$ describe outcomes of a different random experiment in which in each trial one obtains one of two quadruplets with probability ½. For example, after receiving the same $\lambda$ both Alice and Bob flip two fair coins each, and output their outcomes calculated using (13). After N trials Bob sends his N×2 spreadsheet to Alice, who displays her and his results (strictly preserving the order) in a new N×4 spreadsheet. Only these data are described by JP of 4 random variables and now various pair-wise correlations between them can be estimated and they obey strictly (7) for all values of N. This is the main problem in real Bell Tests, because there is no unambiguous ordering between distant clicks produced by entangled photonic signals [15, 33].

It is difficult to understand, why such arguments are not understood and are still a minority stance [24]. Already in 1984, we wrote [52]: *"To describe random events in any particular experiment we do not need to abandon the Kolmogorov axioms of probability theory. However, the measured probabilities in the different experiments may not be determined by conditionalization from a unique probability space. The last assumption was used in all the proofs of Bell inequalities."*

## 6. Conclusions

In LRHVM clicks on detectors (coded ±1) are locally predetermined by variables describing correlated photonic signals, Local predetermination of outcomes of experiments, by some ontic properties of signals, is called usually: *local realism*, *classicality* or *counterfactual definiteness* (CFD). Since different authors attach a different meaning to the notion of *realism*, thus CFD understood as *local predetermination of outcomes* is less ambiguous.

Such assumption was proven incorrect, but it was not stupid. Reinhold Bertlmann remembers, what his friend John said to him: *"I'm a realist… I think that in actual daily practice all scientists are realists, they believe that the world is really there, that it is not a creation of their mind. They feel that there are things there to be discovered, not a world to be invented but a world to be discovered. So I think that realism is a natural position for a scientist and in this debate about the meaning of quantum mechanics I do not know any good arguments against realism."* [53].

This paper is about assumptions underlying various probabilistic hidden variable models and their meaning. CHSH can be rigorously derived for random experiments described by 4 jointly distributed random variables. In EPRB and in Bell Tests such JP does not exist. CFD implies statistical independence, called also *free choice*, *no conspiracy* or MI, and the existence of probabilistic coupling (3–9).

Bell inequalities are violated in several experiments in physics and in cognitive science [54–56], what proves that LRHVM and SHVM provide an incorrect and an oversimplified description of these experiments. Several authors arrived many years ago and often independently, to such correct conclusion [7–12, 14, 15, 26–34, 36–38, 40–52, 57–84], where more references may be found.

In Bell Tests, some data violating inequalities violate also no-signalling [41–45], thus they are also inconsistent with quantum predictions (1, 2) for an ideal EPRB. We demonstrated in [11,15,33] that this apparent violation of no-signalling can be easily explained in a locally causal way, if hidden variables describing measuring instruments are incorporated into probabilistic model.

The speculations about *quantum nonlocality* are based on incorrect interpretations of QM and/or incorrect mental pictures of quantum phenomena [33, 75]. Andrei Khrennikov rejected nonlocality claims using statistical and contextual interpretation of QM [30, 67, 70, 71]. Different arguments were given by Robert Griffiths using consistent histories interpretation of QM [25, 85–89]. These two interpretations differ, but in both interpretations, there is no place for *quantum nonlocality*.

If hidden variables depend on settings, then using Bayes Theorem one concludes that settings depend statistically on hidden variables (12). If *statistical dependence* and *correlation* between distant outcomes are incorrectly interpreted as *causation* one has two options: spooky influences between distant measurements or correlated common causes (possibly coming from a Big Bang) making Nature to comply with the laws of QM.

In QM, it is taken for granted that experimenters can choose their setting as they wish, and Bell never doubted in it. However, after discussions with Shimony, Horn and Clauser [90, 91] he admitted, that the violation of his inequalities might be *explained by superdeterminism* instead of non-local influences. This is what he said in 1985 in BBC interview: *"There is a way to escape the inference of superluminal speeds and spooky action at a distance. But it involves absolute determinism in the universe, the complete absence of free will. Suppose the world is super-deterministic, with not just inanimate nature running on behind-the-scenes clockwork, but with our behavior, including our belief that we are free to choose to do one experiment rather than another, absolutely predetermined, including the "decision" by the experimenter to carry out one set of measurements rather than another, the difficulty disappears. There is no need for a faster than light signal to tell particle A what measurement has been carried out on particle B, because the universe, including particle A, already "knows" what that measurement, and its outcome, will be."*

In 1964, Bell assumed CFD, thus for him it was obvious that hidden variables could not depend on chosen experimental settings. In 1985, hidden variables were assumed to describe any local causes of clicks registered in distant laboratories and *superdeterminism* was recognized as one of possible loopholes in Bell Tests. Jan-Åke Larsson in his excellent review article about possible loopholes in Bell Tests concluded [92]: "The *loophole of superdeterminism cannot be closed by scientific methods; the assumption that the world is not superdeterministic is needed to do science in the first place*". Nevertheless, several toy models explaining how *superdeterminism* allows preserving local causality were proposed [93–95].

Sabine Hossenfelder and Tim Parker succeeded to revive recently discussions about *superdeterminism* [96–98].

In a recent letter to Nature Physics, Jonte Hance and Sabine Hossenfelder [99] correctly insist that: *"the observed violations of Bell's inequality can be said to show that maintaining local causality requires violating statistical independence. We wish to stress that this is not merely an issue of interpretation. The statistical independence assumption is mathematically necessary for the formulation of Bell-type inequalities."* They also correctly underline that *correlation* does not mean *causation* and that *"the mathematical assumption of statistical independence bears no relevance to the philosophical discussion of free will."* We also agree with their conclusion:*"Contrary to what is often stated, these observations do not demonstrate that "spooky action at a distance" is real and nature, therefore, non-local. Rather, the observations show that if nature is local, then statistical independence must be violated."*

However, for Sabine Hossenfelder [97], the violation of *statistical independence* is due to *superdeterminism*, which she defines as: *"Superdeterminism, then, means that the measurement settings are part of what determines the outcome of the time-evolution of the prepared state. What does it mean to violate Statistical Independence? It means that fundamentally everything in the universe is connected with everything else, if subtly so. You may be tempted to ask where these connections come from, but the whole point of superdeterminism is that this is just how nature is. It's one of the fundamental assumptions of the theory, or rather; you could say one drops the usual assumption that such connections are absent."*

In this paper we reviewed and extended the arguments given in [9–12] explaining that if hidden variables describing measuring instruments are correctly incorporated in a probabilistic model the *statistical independence* is violated but neither *retro-causality*, *superderminism* [97] nor *extended causal networks* [100, 101] are needed to explain it. The measuring instruments define a context of a random experiment. Therefore, statistical dependence of hidden variables on the measurement settings should be called *contextuality*.

Correlations between <u>distant</u> experimental outcomes predicted by QM are often called *nonlocal*. It is misleading because using a contextual hidden variable model they may be explained in a local and causal way [11, 15, 33]. Brunner et al. [102] explain that these correlations should be rather called Bell-nonlocal.

We explained that the violation of Bell-CHSH inequalities have quite limited metaphysical implications. Nevertheless, the research stimulated by Bell Theorem and beautiful experiments designed and performed to test inequalities, rewarded recently by a Nobel prize, paved the road to important applications of "nonlocal "quantum correlations in quantum information and in quantum technologies.

For us, the violation of BI-CHSH inequalities in Bell Tests [103–110] proves only that hidden variables have to depend on settings confirming contextual character of quantum observables and an active role played by measuring instruments.

Bell thought that he had to choose between *nonlocality* and *superdeterminism* understood as the *violation of experimenters' freedom of choice*. From two bad choices he chose *nonlocality*. Today he would probably choose *violation of statistical independence* understood as *contextuality*.

**Author Contributions:** The author confirms being the sole contributor of this work.

**Funding:** This research received no external findings.

**Institutional Review Board Statement:** Not applicable.

**Informed Consent Statement:** Not applicable.

**Data Availability Statement:** Not applicable.

**Conflicts of Interest:** The author declares no conflict of interests.